\documentclass[prd,twocolumn,groupedaddress,showpacs,nofootinbib]{revtex4}
\usepackage{graphicx}
\usepackage{dcolumn}
\usepackage{amssymb}
\usepackage{mathrsfs}
\usepackage{amsmath}
\usepackage{epsfig}
\usepackage[dvips]{color}
\usepackage{hhline}

\begin{document}

\title{Multi-component dark matter with magnetic moments for Fermi-LAT gamma-ray line}

\author{Pei-Hong Gu}
\email{peihong.gu@mpi-hd.mpg.de}

\affiliation{Max-Planck-Institut f\"{u}r Kernphysik, Saupfercheckweg
1, 69117 Heidelberg, Germany}

\begin{abstract}

We propose a model of multi-component dark matter with magnetic
moments to explain the $130\,\textrm{GeV}$ gamma-ray line hinted by
the Fermi-LAT data. Specifically, we consider a $U(1)_X^{}$ dark
sector which contains two vector-like fermions besides the related
gauge and Higgs fields. A very heavy messenger scalar is further
introduced to construct the Yukawa couplings of the dark fermions to
the heavy $[SU(2)^{}]$-singlet leptons in the $SU(3)_c^{}\times
SU(2)_L^{}\times SU(2)_R^{}\times U(1)_{B-L}^{}$ left-right
symmetric models for universal seesaw. A heavier dark fermion with a
very long lifetime can mostly decay into a lighter dark fermion and
a photon at one-loop level. The dark fermions can serve as the dark
matter particles benefited from their annihilations into the dark
gauge and Higgs fields. In the presence of a $U(1)$ kinetic mixing,
the dark matter fermions can be verified by the ongoing and
forthcoming dark matter direct detection experiments.

\end{abstract}


\maketitle

\section{Introduction}

Astronomical and cosmological observations indicate the existence of
dark matter in the present universe. Many dark matter candidates
have been suggested in various scenarios beyond the
$SU(3)_c^{}\times SU(2)_L^{}\times U(1)_{Y}^{}$ standard model (SM).
The dark matter particles can be directly detected through their
scatterings off the familiar nucleons and/or indirectly detected
through their annihilations/decays into the SM species. Recently,
the Fermi-LAT data on the cosmic gamma-ray spectrum from the
Galactic center (GC) have revealed a tentative evidence for a
line-like feature at an energy around $130\,\textrm{GeV}$
\cite{bhivw2012,weniger2012,thr2012,gk2012,sf2012,ly2012,clsw2012,ctu2012,yyffc2012,
hrt2012,bbcfw2012,hyybc2012,hl2012,whiteson2012,hrt2012-2,fsw2012}.
Such monochromatic photon can be induced by a dark matter
annihilation or decay. There have been a lot of models
\cite{dmpr2012,cline2012,cs2012,rtw2012,kp2012,lpp2012,bh2012,chst2012,
dem2012,klll2012,bg2012,pp2012,fhkms2012,tyz2012,cf2012,bs2012,kgpmm2012,lndh2012,
bergstrom2012,wh2012,lpp2012-2,fr2012,wy2012,shakya2012,bks2012,chp2012,dmt2012,ssw2012,
yr2012,cdm2012,rtw2012-2,bsz2012,zhang2012,bbes2012,lps2012,knpz2013,bmsb2013}
realizing the required dark matter annihilations or decays.

Usually, we need some particles heavier than the dark matter to
mediate the significant annihilations or decays of the dark matter
into the monochromatic photons at loop level since the dark matter
is only allowed to have an extremely tiny electric charge
\cite{myz2011}. On the other hand, in the $SU(3)_c^{}\times
SU(2)_L^{}\times SU(2)_R^{}\times U(1)_{B-L}^{}$ left-right
symmetric models \cite{ps1974} for universal \cite{berezhiani1983}
seesaw \cite{minkowski1977,mw1980}, some heavy $[SU(2)^{}]$-singlet
leptons and quarks are well motivated to alleviate the hierarchy
among the SM fermion masses and solve the strong CP problem
\cite{bm1989}. Such heavy fermions could take part in mediating the
radiative dark matter annihilations or decays for the
$130\,\textrm{GeV}$ gamma-ray line. In this paper, we shall
demonstrate this possibility in a multi-component dark matter
\cite{bk1991,bfs2003,ma2006,gu2007,hln2007,cmw2007,hambye2008,ghsz2009,adkt2012,hz2012}
scenario. Specifically, we shall introduce a $U(1)_X^{}$ dark sector
which contains two vector-like fermions besides the related gauge
and Higgs fields. There is also a very heavy messenger scalar having
the Yukawa couplings with the dark fermions and the heavy leptons.
The heavier dark fermion can have a very long lifetime although it
mostly decays into the lighter dark fermion with a photon. The dark
fermions can obtain the desired dark matter relic density through
their annihilations into the dark gauge and Higgs fields. As the
$U(1)_X^{}$ and $U(1)_{B-L}^{}$ gauge fields are allowed to have a
kinetic mixing, the dark matter fermions can be verified by the
ongoing and forthcoming dark matter direct detection experiments.

\section{The model}

For simplicity, we will not give the full Lagrangian. Instead, we
only show the kinetic, mass and Yukawa terms relevant to our
demonstration,
\begin{eqnarray}
\label{lagrangian}
\mathcal{L}&\supset&-\frac{1}{4}W_{L\mu\nu}^{a}W^{a}_{L\mu\nu}-\frac{1}{4}W_{R\mu\nu}^{a}W^{a}_{R\mu\nu}
-\frac{1}{4}B_{\mu\nu}^{}B^{}_{\mu\nu}\nonumber\\
&&-\frac{1}{4}C_{\mu\nu}^{}C^{\mu\nu}_{}
-\frac{\epsilon}{2}B_{\mu\nu}^{}C^{\mu\nu}_{}+(D_\mu^{}\phi_{L}^{})^\dagger_{}D^\mu_{}\phi_{L}^{}\nonumber\\
&&
+(D_\mu^{}\phi_{R}^{})^\dagger_{}D^\mu_{}\phi_{R}^{}+(D_\mu^{}\sigma)^\dagger_{}D^\mu
\sigma+(D_\mu^{}\delta)^\dagger_{}D^\mu
\delta\nonumber\\
&&+i\bar{q}^{}_{L}\!\not\!\!D q^{}_{L}+i\bar{q}^{}_{R}\!\not\!\!D
q^{}_{R}+i\bar{l}^{}_{L}\!\not\!\!D
l^{}_{L}+i\bar{l}^{}_{R}\!\not\!\!D
l^{}_{R}\nonumber\\
&&+i\overline{D}^{}_{L}\!\not\!\!D
D^{}_{L}+i\overline{D}^{}_{R}\!\not\!\!D
D^{}_{R}+i\overline{U}^{}_{L}\!\not\!\!D
U^{}_{L}+i\overline{U}^{}_{R}\!\not\!\!D
U^{}_{R}\nonumber\\
&&+i\overline{E}^{}_{L}\!\not\!\!D
E^{}_{L}+i\overline{E}^{}_{R}\!\not\!\!D
E^{}_{R}+i\bar{\chi}_{L}^{}\!\not\!\!D\chi_{L}^{}
+i\bar{\chi}_{R}^{}\!\not\!\!D\chi_{R}^{}\nonumber\\
&&-M_\delta^2\delta^\dagger_{}\delta
-\overline{D}_{L}^{}M_{D}^{}D_{R}^{}
-\overline{U}_{L}^{}M_{U}^{}U_{R}^{}-\overline{E}_{L}^{}M_{E}^{}E_{R}^{}\nonumber\\
&&
-\bar{\chi}_{L}^{}m_{\chi}^{}\chi_{R}^{}
-\bar{q}_{L}^{}\phi_L^{}y_D^{L}D_{R}^{}-\bar{q}_{R}^{}\phi_R^{}y^R_D D_{L}^{}\nonumber\\
&&-\bar{q}_{L}^{}\tilde{\phi}_L^{}y_U^LU_{R}^{}-\bar{q}_{R}^{}\tilde{\phi}_R^{}y_U^R
U_{L}^{}
-\bar{l}_{L}^{}\phi_L^{}y_E^L E_{R}^{}\nonumber\\
&&-\bar{l}_{R}^{}\phi_R^{}y_E^R E_{L}^{}
-\delta\overline{E}_{L}^{}f_R^{}\chi_{R}^{}
-\delta\overline{E}_{R}^{}f_L^{}\chi_{L}^{}+\textrm{H.c.}\,.
\end{eqnarray}
Here $W_{L}^a$, $W_{R}^a$, $B$ and $C$ are the gauge fields
associated with the $SU(2)_{L}^{}$, $SU(2)_R^{}$, $U(1)_{B-L}^{}$
and $U(1)_X^{}$ gauge groups, respectively. The Higgs scalars
\begin{eqnarray}
\phi_L^{}(+1,0)=\left[\begin{array}{c}\phi_L^{+}\\
[1mm]
\phi_L^{0}\end{array}\right]\,,~\phi_R^{}(+1,0)=\left[\begin{array}{c}\phi_R^{+}\\
[1mm] \phi_R^{0}\end{array}\right]\,, ~\sigma(0,+2)
\end{eqnarray}
are an $SU(2)_L^{}$ doublet, an $SU(2)_R^{}$ doublet and an $SU(2)$
singlet, respectively. Here and thereafter the first and second
numbers in parentheses are the $U(1)^{}_{B-L}$ charges $B-L$ and the
$U(1)_X^{}$ charges $X$. The messenger scalar $\delta$ is an $SU(2)$
singlet and carries both of the $U(1)^{}_{B-L}$ and $U(1)_X^{}$
charges,
\begin{eqnarray}
\delta(-2,+\frac{2}{3})\,.
\end{eqnarray}
Among the fermions, the $[SU(2)]$-doublet quarks $q_{L,R}^{}$, the
$[SU(2)]$-doublet leptons $l_{L,R}^{}$, the $[SU(2)]$-singlet quarks
$D_{L,R}^{}$ and $U_{L,R}^{}$ as well as the $[SU(2)]$-singlet
leptons $E_{L,R}^{}$ carry the $U(1)^{}_{B-L}$ charges,
\begin{eqnarray}
&&q_{L}^{}(+\frac{1}{3},0)=\left[\begin{array}{c}u_{L}^{}\\
[1mm]
d_{L}^{}\end{array}\right]\,,~~q_{R}^{}(+\frac{1}{3},0)=\left[\begin{array}{c}u_{R}^{}\\
[1mm]
d_{R}^{}\end{array}\right]\,,\nonumber\\
[2mm]
&&~l_{L}^{}(-1,0)=\left[\begin{array}{c}\nu_{L}^{}\\
[1mm]
e_{L}^{}\end{array}\right]\,,~~~~l_{R}^{}(-1,0)=\left[\begin{array}{c}\nu_{R}^{}\\
[1mm]
e_{R}^{}\end{array}\right]\,,\nonumber\\
[2mm]
&&D_{L,R}^{}(-\frac{2}{3},0)\,,~~U_{L,R}^{}(+\frac{4}{3},0)\,,~~E_{L,R}^{}(-2,0)\,,
\end{eqnarray}
while the $[SU(2)]$-singlet dark fermions $\chi_{L,R}^{}$ carry the
$U(1)_X^{}$ charges,
\begin{eqnarray}
\chi_{L,R}^{}(0,-\frac{2}{3})\,.
\end{eqnarray}
The covariant derivatives are
\begin{eqnarray}
\label{covariant} D_\mu^{}&=&\partial_\mu^{}-i g_{X}^{}\frac{X}{2}
C_\mu^{}-i g_{B-L}^{}\frac{B-L}{2} B_\mu^{}-ig I_{3L}^{} W_{L\mu}^3
\nonumber\\
&&-ig_R^{}I_{3R}^{}W_{R\mu}^3 +...\,,
\end{eqnarray}
where we have only written down the diagonal components of the
$SU(2)$ gauge fields $(W^3_L,\,W^3_R)$ which can mix with the other
$U(1)$ gauge fields $(B,\,C)$.

\subsection{Symmetry breaking}

The $[SU(2)]$-doublet Higgs scalars $\phi_{R}^{}$ and $\phi_{L}^{}$
are responsible for breaking the left-right symmetry down to the
electroweak symmetry and then the electromagnetic symmetry, i.e.
\begin{eqnarray}
&SU(2)_L^{}\times SU(2)_R^{}\times U(1)_{B-L}^{}&\nonumber\\
&\downarrow \langle\phi_{R}^{}\rangle&\nonumber\\
& SU(2)_L^{}\times
U(1)_{Y}^{}&\nonumber\\
&\downarrow\langle\phi_{L}^{}\rangle&\nonumber\\
& U(1)_{em}^{}\,,&
\end{eqnarray}
where $\langle\phi_{R}^{}\rangle$ and $\langle\phi_{L}^{}\rangle$
are the vacuum expectation values (VEVs),
\begin{eqnarray}
\langle\phi_{L}^{}\rangle&=&\langle\phi_{L}^{0}\rangle
=\frac{1}{\sqrt{2}}v_{L}^{}~~(v_L^{}\simeq 246\,\textrm{GeV})\,,\nonumber\\
\langle\phi_{R}^{}\rangle&=&\langle\phi_{R}^{0}\rangle
=\frac{1}{\sqrt{2}}v_{R}^{}\,.
\end{eqnarray}
As for the dark symmetry $U(1)_X^{}$, it will be broken when the
$[SU(2)]$-singlet Higgs scalar $\sigma$ develops its VEV,
\begin{eqnarray}
\langle\sigma\rangle =\frac{1}{\sqrt{2}}v_{X}^{}\,.
\end{eqnarray}
Roughly, the above symmetry breakings take place at the temperatures
$T=\mathcal{O}(\langle\phi_{R}^{}\rangle)$,
$\mathcal{O}(\langle\phi_{L}^{}\rangle)$ and
$\mathcal{O}(\langle\sigma\rangle)$, respectively.

\subsection{Fermions}

The charged fermions have the masses,
\begin{eqnarray}
\mathcal{L}&\supset& -[\bar{d}^{}_L~
\overline{D}^{}_L]\left[\begin{array}{cc}0&\frac{1}{\sqrt{2}}y_{D}^{L}v_L^{}\\
[2mm]
\frac{1}{\sqrt{2}}y^{R\dagger}_D v_R^{}&M_D^{}\end{array}\right]\left[\begin{array}{c}d^{}_R\\
[2mm] D^{}_R\end{array}\right]\nonumber\\
[2mm] && -[\bar{u}^{}_L~
\overline{U}^{}_L]\left[\begin{array}{cc}0&\frac{1}{\sqrt{2}}y_{U}^{L}v_L^{}\\
[2mm]
\frac{1}{\sqrt{2}}y^{R\dagger}_U v_R^{}&M_U^{}\end{array}\right]\left[\begin{array}{c}u^{}_R\\
[2mm] U^{}_R\end{array}\right]\nonumber\\
[2mm] && -[\bar{e}^{}_L~
\overline{E}^{}_L]\left[\begin{array}{cc}0&\frac{1}{\sqrt{2}}y^{L}_{E}v_L^{}\\
[2mm]
\frac{1}{\sqrt{2}}y^{R\dagger}_E v_R^{}&M_E^{}\end{array}\right]\left[\begin{array}{c}e^{}_R\\
[2mm] E^{}_R\end{array}\right]+\textrm{H.c.}\,,\nonumber\\
&&
\end{eqnarray}
which can be block diagonalized by
\begin{eqnarray}
\mathcal{L}&\supset&
-\bar{d}_L^{}\left(-y_D^{L}\frac{v_L^{}v_R^{}}{2M_D^{}}y_D^{R\dagger}\right)
d_R^{}-\overline{D}_L^{}M_D^{}D_R^{}\nonumber\\
&&-\bar{u}_L^{}\left(-y_U^{L}\frac{v_L^{}v_R^{}}{2M_U^{}}y_U^{R\dagger}\right)
u_R^{}-\overline{U}_L^{}M_U^{}U_R^{}\nonumber\\
&&
-\bar{e}_L^{}\left(-y_E^{L}\frac{v_L^{}v_R^{}}{2M_E^{}}y^{R\dagger}_E\right)
e_R^{}-\overline{E}_L^{}M_E^{}E_R^{}\nonumber\\
&&+\textrm{H.c.}\,,
\end{eqnarray}
Remarkably, we have the heavy charged fermions besides the SM
charged fermions in this universal seesaw scenario.

In the following we will work in the base where the mass matrices of
the dark fermions $\chi$ and the heavy charged fermions $F=D,U,E$
are diagonal and real, i.e.
\begin{eqnarray}
m_\chi^{}&=&\textrm{diag}\{m_{\chi_1^{}}^{}\,,~m_{\chi_2^{}}^{}\}\,,\nonumber\\
M_F^{}&=&\textrm{diag}\{M_{F_1^{}}^{}\,,~M_{F_2^{}}^{}\,,~M_{F_3^{}}^{}\}\,,
\end{eqnarray}
and then define the vector-like fermions:
\begin{eqnarray}
\chi_i^{}=\chi_{Li}^{}+\chi_{Ri}^{}\,,~~F_a^{}=F_{La}^{}+F_{Ra}^{}\,.
\end{eqnarray}
Moreover, we will assume $m_{\chi_1^{}}^{}\leq m_{\chi_2^{}}^{}$ and
$M_{F_1^{}}^{}\leq M_{F_2^{}}^{}\leq M_{F_3^{}}^{}$ without loss of
generality. Note the left-right symmetry breaking scale and the
heavy fermion masses should be large enough to escape from the
experimental constraint. For example, the right-handed charged gauge
boson should be heavier than a few TeV \cite{zajm2007}. In the
present work, we will take $v_R^{}\sim
M_{F_{1,2,3}^{}}^{}=\mathcal{O}(10^3_{}\,\textrm{TeV})$ to give a
numerical example.

The present model can accommodate a discrete parity symmetry to
solve the strong CP problem without an axion \cite{bm1989,bcs1991}.
Furthermore, the neutral neutrinos can have a two-loop induced Dirac
mass matrix proportional to the SM charged lepton mass matrix
\cite{mohapatra1988}. One can introduce more fermions or scalars to
generate the desired neutrino masses and mixing.

\subsection{Gauge fields}

We can remove the kinetic mixing between the $U(1)^{}_X$ and
$U(1)^{}_{B-L}$ gauge fields by making a non-unitary transformation
\cite{fh1991},
\begin{eqnarray}
B_\mu^{}&=&\tilde{B}_\mu^{}-\frac{\epsilon}{\sqrt{1-\epsilon^2_{}}}\tilde{C}_\mu^{}=\tilde{B}_\mu^{}-\xi
\tilde{C}_\mu^{} \,,\nonumber\\
C_\mu^{}&=&\frac{1}{\sqrt{1-\epsilon^2}}\tilde{C}_\mu^{}\,,
\end{eqnarray}
and then define the orthogonal fields,
\begin{subequations}
\label{orthogonal}
\begin{eqnarray}
A_\mu^{}&=&W^3_{L\mu}s_W^{}
+(W^3_{R\mu}s_R^{}+\tilde{B}_{\mu}^{}c_R^{})c_W^{}\,,\\
Z_{L\mu}^{}&=&W^3_{L\mu}c_W^{}
-(W^3_{R\mu}s_R^{}+\tilde{B}_{\mu}^{}c_R^{})s_W^{}\,,\\
Z_{R\mu}^{}&=&W^3_{R\mu}c_R^{}-\tilde{B}_{\mu}^{}s_R^{}\,,\\
Z_{X\mu}^{}&=&\tilde{C}_{\mu}^{}\,,
\end{eqnarray}
\end{subequations}
with
\begin{eqnarray}
&&s_R^{}=\sin\theta_R^{}\,,
~~c_R^{}=\cos\theta_R^{}~~\textrm{for}~~t_R^{}=\tan\theta_R^{}=\frac{g_{B-L}^{}}{g_R^{}}\,,
\nonumber\\
&&
s_W^{}=\sin\theta_W^{}\,,~~c_W^{}=\cos\theta_W^{}~~\textrm{for}\nonumber\\
&& \quad\quad\quad\quad
t_W^{}=\tan\theta_W^{}=\frac{g_{B-L}^{}g_{R}^{}/g}{\sqrt{g_{B-L}^{2}+g_R^2}}=\frac{g'}{g}\,.
\end{eqnarray}
Here $g$ and $g'$ are the SM gauge couplings with $g\simeq 0.653$
and $g'\simeq 0.358$ while $\theta_W^{}$ is the Weinberg angle
$s_W^2\simeq 0.231$. If a parity symmetry is imposed, we can
determine the unknown gauge couplings $g_R^{}$ and $g_{B-L}^{}$ by
\begin{eqnarray}
g_R^{}=g\simeq
0.653\,,~~g_{B-L}^{}=\frac{gg'}{\sqrt{g^2_{}-g'^2_{}}}\simeq
0.428\,.
\end{eqnarray}
Among the orthogonal fields (\ref{orthogonal}), $A$ is the massless
photon $\gamma$, while $Z_{L}^{}$, $Z_{R}^{}$ and $Z_{X}^{}$ have
the mass terms as below,
\begin{eqnarray}
\mathcal{L}&\supset&\frac{1}{2}m_{Z_R^{}}^2 \left(Z_{R\mu}^{}+\xi
Z_{X\mu}^{}s_R^{}\right)
\left(Z_{R}^{\mu}+\xi Z_{X}^{\mu}s_R^{}\right)\nonumber\\
&&+\frac{1}{2}m_{Z_L^{}}^2
\left(Z_{L\mu}^{}+Z_{R\mu}^{}t_R^{}s_W^{}+\xi Z_{X\mu}^{}t_R^{}s_W^{}\right)\nonumber\\
&&\times \left(Z_{L}^{\mu}+Z_{R}^{\mu}t_R^{}s_W^{}+\xi
Z_{X}^{\mu}t_R^{}s_W^{}\right)+\frac{1}{2}m_{Z_X^{}}^2
Z_{X\mu}^{}Z_{X}^{\mu}\nonumber\\
&&~~\textrm{with}~~\left\{\begin{array}{ccl}
m_{Z_R^{}}^{}&=&\frac{1}{2\cos\theta_R^{}}g_R^{}v_R^{}\,,\\
[2mm]
m_{Z_L^{}}^{}&=&\frac{1}{2\cos\theta_W^{}}gv_L^{}\simeq 91\,\textrm{GeV}\,,\\
[2mm]
m_{Z_X^{}}^{}&=&\frac{1}{\sqrt{1-\epsilon^2}}g_X^{}v_X^{}\,.\end{array}\right.
\end{eqnarray}

In the following we will focus on the case that the orthogonal
fields $Z_{R}^{}$, $Z_{L}^{}$ and $Z_{X}^{}$ approximate to the mass
eigenstates for $m_{Z_R^{}}^{2}\gg m_{Z_L^{}}^{2}\gg m_{Z_X^{}}^{2}$
and $\epsilon\ll 1$. In this case, the quasi-mass-eigenstate
$Z_L^{}$ is identified to the SM $Z$ boson.

\section{Dark gauge boson decay}

The dark gauge filed $C$ which is mostly the quasi-mass-eigenstate
$Z_X^{}$ can couple to the SM fermions besides the dark fermions,
\begin{eqnarray}
\mathcal{L}&\supset& -\frac{1}{6}\xi g_{B-L}^{}\bar{d}\gamma^\mu_{}d
Z_{X\mu}^{}-\frac{1}{6}\xi g_{B-L}^{}\bar{u}\gamma^\mu_{}u
Z_{X\mu}^{}\nonumber\\
&&+\frac{1}{2}\xi g_{B-L}^{}\bar{e}\gamma^\mu_{}e
Z_{X\mu}^{}+\frac{1}{2}\xi g_{B-L}^{}\bar{\nu}\gamma^\mu_{}\nu
Z_{X\mu}^{}\nonumber\\
&&-\frac{1}{3\sqrt{1-\epsilon^2_{}}}g_X^{}\bar{\chi}\gamma^\mu_{}\chi
Z_{X\mu}^{}\,.
\end{eqnarray}
Therefore, if the dark gauge boson $Z_X^{}$ is heavy enough, it can
decay into the SM fermion pairs $f\bar{f}$. For
$2m_f^{}<m_{Z_X^{}}^{}< 2m_{\chi_1^{}}^{}$, the decay width should
be
\begin{eqnarray}
\Gamma_{Z_X^{}}^{}&=&\sum_f^{}\Gamma_{Z_X^{}\rightarrow
f\bar{f}}^{}\nonumber\\
&\simeq&\sum_{f}^{}N_c^{}\frac{\epsilon^2_{}g_{B-L}^2}{12\pi}\left(\frac{B-L}{2}\right)^2_{}
m_{Z_X^{}}^{}\left(1+\frac{m_f^2}{m_{Z_X^{}}^2}\right)\nonumber\\
&&\times\sqrt{1-4\frac{m_f^2}{m_{Z_X^{}}^2}}\,,
\end{eqnarray}
where $(N_c^{},B-L)=(3,\frac{1}{3})$ for a quark and
$(N_c^{},B-L)=(1,-1)$ for a lepton. We then find
\begin{eqnarray}
\tau_{Z_X^{}}^{}\simeq \left(\frac{1.34\times
10^{-11}_{}}{\epsilon}\right)^2_{}
\left(\frac{0.428}{g_{B-L}^{}}\right)^2_{}\left(\frac{500\,\textrm{MeV}}{m_{Z_X^{}}^{}}\right)\textrm{sec}\,.
\end{eqnarray}
So, the dark gauge boson $Z_X^{}$ with a mass
$m_{Z_X^{}}^{}=500\,\textrm{MeV}$ can have a lifetime shorter than
$1$ second if we take $\epsilon> 1.34\times 10^{-11}_{}$. Currently,
the measurement on the muon magnetic moment constrains $\epsilon^2
c_R^2 c_W^2< 2\times 10^{-4}_{}$ for
$m_{Z_X^{}}^{}=500\,\textrm{MeV}$ \cite{pospelov2008}.

\section{Dark matter relic density}

\begin{figure*}
\vspace{6.0cm} \epsfig{file=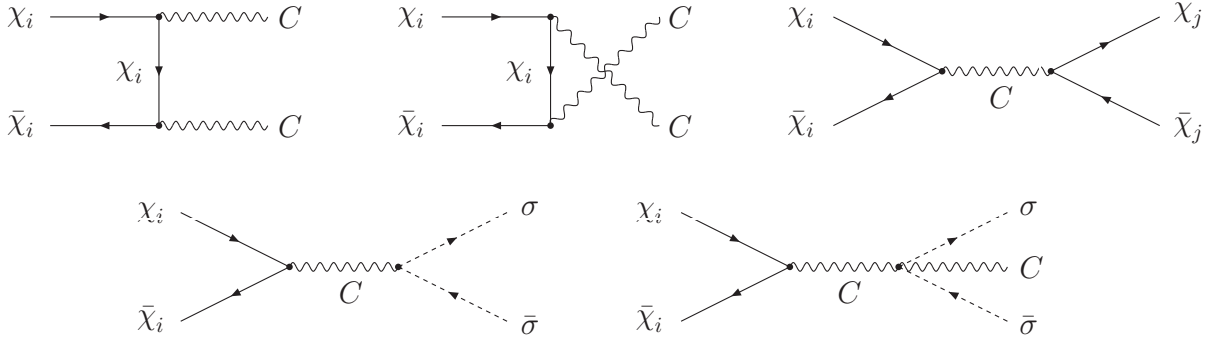, bbllx=5.9cm, bblly=6.0cm,
bburx=15.9cm, bbury=16cm, width=8.2cm, height=8.2cm, angle=0,
clip=0} \vspace{-8.3cm} \caption{\label{annihilation} The dark
fermion annihilations. Here $\chi_{i,j}^{}~(i\neq j)$ denotes the
dark matter fermions, $C$ is the $U(1)_X^{}$ gauge field, while
$\sigma$ is the Higgs scalar for breaking the $U(1)_X^{}$ symmetry.}
\end{figure*}

The dark fermions $\chi_{1,2}^{}$ can annihilate into the dark gauge
field $C$ and the dark Higgs field $\sigma$. The heavier dark
fermion $\chi_{2}^{}$ can also annihilate into the lighter dark
fermion $\chi_{1}^{}$. The relevant diagrams are shown in Fig.
\ref{annihilation}. If the annihilations freeze out before the
$U(1)_X^{}$ symmetry breaking at the temperature
$T=\mathcal{O}(\langle\sigma\rangle)$, the annihilation cross
sections should be dominated by
\begin{eqnarray}
\sigma_{\chi_1^{}}^{}&=&\langle\sigma_{\chi_1^{}\bar{\chi}_1^{}\rightarrow
CC}^{}v_{\textrm{rel}}^{}\rangle +
\langle\sigma_{\chi_1^{}\bar{\chi}_1^{}\rightarrow
\sigma\bar{\sigma}}^{}v_{\textrm{rel}}^{}\rangle \nonumber\\
&\simeq&
\frac{13}{5184\pi}\frac{g_X^4}{m_{\chi_1^{}}^2}\,,\\
\sigma_{\chi_2^{}}^{}&=&\langle\sigma_{\chi_2^{}\bar{\chi}_2^{}\rightarrow
C C}^{}v_{\textrm{rel}}^{}\rangle+
\langle\sigma_{\chi_2^{}\bar{\chi}_2^{}\rightarrow
\sigma\bar{\sigma}}^{}v_{\textrm{rel}}^{}\rangle\nonumber\\
&& +\langle\sigma_{\chi_2^{}\bar{\chi}_2^{}\rightarrow
\chi_1^{}\bar{\chi}_1^{}}^{}v_{\textrm{rel}}^{}\rangle\nonumber\\
&\simeq& \frac{13}{5184\pi}\frac{g_X^4}{m_{\chi_2^{}}^2}
\left[1+\frac{4}{13}\left(1+\frac{m_{\chi_1^{}}^2}{2m_{\chi_2^{}}^2}\right)
\sqrt{1-\frac{m_{\chi_1^{}}^2}{m_{\chi_2^{}}^2}}\right]\,.\nonumber\\
&&
\end{eqnarray}
Here $v_{\textrm{rel}}^{}$ is the relative velocity between the two
annihilating particles in their center-of-mass frame.

As we will clarify later, the heavier dark fermion $\chi_2^{}$ has a
very long lifetime and hence contributes to the dark matter relic
density together with the lighter and stable dark fermion
$\chi_{1}^{}$. The relic density of the dark fermion $\chi_{i}^{}$
can be calculated by \cite{kt1990}
\begin{eqnarray}
\Omega_{\chi_i^{}+\bar{\chi}_i^{}}^{}h^2_{}=\Omega_{\chi_i^{}}^{}h^2_{}+\Omega_{\bar{\chi}_i^{}}^{}h^2_{}=\frac{1.07\times
10^{9}_{}m_{\chi^{}_{i}}^{}}{\sqrt{g_\ast^{}}M_{\textrm{Pl}}^{}\sigma_{\chi_i^{}}^{}T_i^{}(\textrm{GeV})}\,,
\end{eqnarray}
where $M_{\textrm{Pl}}^{}=1.22\times 10^{19}_{}\,\textrm{GeV}$ is
the Planck mass, $T_i^{}$ is the freeze-out temperature determined
by \cite{kt1990}
\begin{eqnarray}
\frac{m_{\chi^{}_{i}}^{}}{T_i^{}}&=& \ln(2\times 0.038\,
M_{\textrm{Pl}}^{}m_{\chi^{}_{i}}^{}\sigma_{\chi_i^{}}^{}/\sqrt{g_\ast^{}})\nonumber\\
&&-\frac{1}{2}\ln[\ln(2\times 0.038\,
M_{\textrm{Pl}}^{}m_{\chi^{}_{i}}^{}\sigma_{\chi_i^{}}^{}/\sqrt{g_\ast^{}})]\,,~~~~~
\end{eqnarray}
while $g_{\ast}^{}=g_{\ast}^{}(T_i^{})$ is the number of
relativistic degrees of freedom. The dark fermions $\chi_{1,2}^{}$
can dominate the dark matter relic density for a proper parameter
choice. For example, we input
\begin{eqnarray}
&&g_X^{}=0.592\,,~v_X^{}=845\,\textrm{MeV}\,,~
m_{\chi_1^{}}^{}=20\,\textrm{GeV}\,,\nonumber\\
&&m_{\chi_2^{}}^{}=262\,\textrm{GeV}\,,~g_\ast^{}(T_1^{})=84\,,~g_\ast^{}(T_2^{})=94\,,
\end{eqnarray}
to obtain
\begin{eqnarray}
&&\sigma_{\chi_1^{}}^{}\simeq 2.45\times
10^{-7}_{}\,\textrm{GeV}^{-2}_{}\,,~~T_1^{}\simeq
m_{\chi_1^{}}^{}/25.3\,,\nonumber\\
&&\Omega_{\chi_{1}^{}+\bar{\chi}_1^{}}^{}h^2_{}\simeq 0.001\,,
\end{eqnarray}
and
\begin{eqnarray}
&&\sigma_{\chi_2^{}}^{}\simeq 1.87\times
10^{-9}_{}\,\textrm{GeV}^{-2}_{}\,,~~T_2^{}\simeq
m_{\chi_2^{}}^{}/23.0\,,\nonumber\\
&&\Omega_{\chi_{2}^{}+\bar{\chi}_2^{}}^{}h^2_{}\simeq 0.111\,.
\end{eqnarray}
The total dark matter relic density thus should be
\begin{eqnarray}
\Omega_{\textrm{DM}}^{}h^2_{}=\Omega_{\chi_{1}^{}+\bar{\chi}_1^{}}^{}h^2_{}
+\Omega_{\chi_{2}^{}+\bar{\chi}_2^{}}^{}h^2_{}\simeq 0.112\,,
\end{eqnarray}
which is well consistent with the observations \cite{larson2010}. In
the above numerical estimation, the heavier dark fermion $\chi_2^{}$
rather than the lighter dark fermion $\chi_1^{}$ dominates the dark
matter relic density, i.e.
\begin{eqnarray}
\frac{\Omega_{\chi_{1}^{}+\bar{\chi}_1^{}}^{}h^2_{}}{\Omega_{\textrm{DM}}^{}h^2_{}}\simeq
0.9\%\,,~~ \frac{\Omega_{\chi_{2}^{}+\bar{\chi}_2^{}}^{}h^2_{}}
{\Omega_{\textrm{DM}}^{}h^2_{}}\simeq99.1\%\,,
\end{eqnarray}
as a result of their hierarchial masses $m_{\chi_2^{}}^2\gg
m_{\chi_1^{}}^{2}$. Actually, it is easy to see
\begin{eqnarray}
\frac{\Omega_{\chi_{1}^{}+\bar{\chi}_1^{}}^{}h^2_{}}{\Omega_{\textrm{DM}}^{}h^2_{}}\leq
50\%\leq
\frac{\Omega_{\chi_{2}^{}+\bar{\chi}_2^{}}^{}h^2_{}}{\Omega_{\textrm{DM}}^{}h^2_{}}
~~\textrm{for}~~ m_{\chi_1^{}}^{}\leq m_{\chi_2^{}}^{}\,.
\end{eqnarray}

\section{Dark matter decay}

\begin{figure*}
\vspace{7.0cm} \epsfig{file=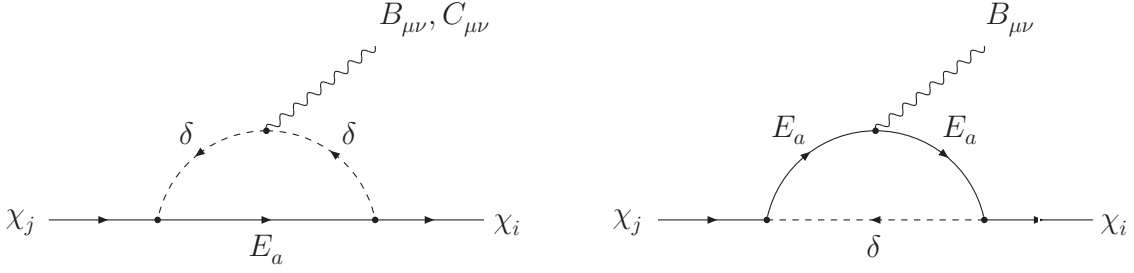, bbllx=5.5cm, bblly=6.0cm,
bburx=15.5cm, bbury=16cm, width=8.2cm, height=8.2cm, angle=0,
clip=0} \vspace{-10.7cm} \caption{\label{magnetic} The dark matter
fermions $\chi_{i,j}^{}$ couple to the $U(1)_{B-L}^{}$ and
$U(1)_X^{}$ field strength tensors ($B_{\mu\nu}^{}, C_{\mu\nu}^{}$)
at one-loop level. }
\end{figure*}

The dark matter fermions $\chi_{1,2}^{}$ can couple to the
$U(1)_{B-L}^{}$ and $U(1)_X^{}$ field strength tensors
($B_{\mu\nu}^{}, C_{\mu\nu}^{}$) at one-loop level. We show the
relevant diagrams in Fig. \ref{magnetic}. For
$m_{\chi_{1,2}^{}}^{}\ll M_{E_a^{}}^{},M_{\delta}^{}$, the effective
interactions should be
\begin{eqnarray}
\mathcal{L}&\supset&-\lambda_{ij}^{}\bar{\chi}_{Li^{}}^{}\sigma^{\mu\nu}_{}\chi_{Rj}^{}B_{\mu\nu}^{}
-\kappa_{ij}^{}\bar{\chi}_{Li^{}}^{}\sigma^{\mu\nu}_{}\chi_{Rj}^{}C_{\mu\nu}^{}+\textrm{H.c.}\nonumber\\
&=&-\lambda_{ij}^{}\bar{\chi}_{Li^{}}^{}\sigma^{\mu\nu}_{}\chi_{Rj}^{}(A_{\mu\nu}^{}c_W^{}c_R^{}
-Z_{L\mu\nu}^{}s_W^{}c_R^{}\nonumber\\
&& -Z_{R\mu\nu}^{}s_R^{}-\xi Z_{X\mu\nu}^{})
-\frac{1}{\sqrt{1-\epsilon^2_{}}}\kappa_{ij}^{}\bar{\chi}_{Li^{}}^{}\sigma^{\mu\nu}_{}\chi_{Rj}^{}Z_{X\mu\nu}^{}
\nonumber\\
&&+\textrm{H.c.}\,,
\end{eqnarray}
where the couplings $\lambda_{ij}^{}$ and $\kappa_{ij}^{}$ are given
by
\begin{eqnarray}
\lambda_{ij}^{}&=&\frac{g_{B-L}^{}}{32\pi^2_{}}(f_L^\dagger)_{ia}^{}
(f_R^{})_{aj}^{}\frac{1}{M_{E_a^{}}}F_B^{}\left(\frac{M_{\delta}^2}{M_{E_a^{}}^2}\right)
~~\textrm{with}\nonumber\\
&&F_B^{}(x)=\frac{1}{1-x} +\frac{x}{(1-x)^2_{}}\ln x\,,\\
\kappa_{ij}^{}&=&\frac{g_{X}^{}}{96\pi^2_{}}(f_L^\dagger)_{ia}^{}
(f_R^{})_{aj}^{}\frac{1}{M_{E_a^{}}}F_C^{}\left(\frac{M_{\delta}^2}{M_{E_a^{}}^2}\right)
~~\textrm{with}\nonumber\\
&&F_C^{}(x)=-\frac{1+x}{2(1-x)^2_{}} -\frac{x}{(1-x)^3_{}}\ln x\,.
\end{eqnarray}
Therefore, the heavier dark matter fermion $\chi_2^{}$ can decay
into the lighter dark matter fermion $\chi_1^{}$ and a gauge boson
($\gamma$, $Z$, $Z_X^{}$ or $Z_R^{}$) as long as the kinematics is
allowed. Provided that
\begin{eqnarray}
m_{Z_R^{}}^{}\gg m_{\chi_2^{}}^{}- m_{\chi_1^{}}^{}>
m_{Z}^{}\,,~m_{Z_X^{}}^{}\,,
\end{eqnarray}
we can obtain the decay widths,
\begin{eqnarray}
\Gamma_{\chi_2^{}\rightarrow
\chi_1^{}\gamma}^{}&=&\Gamma_{\bar{\chi}_2^{}\rightarrow
\bar{\chi}_1^{}\gamma}^{}\nonumber\\
&\simeq& \frac{c_R^{2}c_W^{2}}{2\pi}|\lambda_{12}^{}|^2_{}
m_{\chi_2^{}}^3(1-m_{\chi_1^{}}^2/m_{\chi_2^{}}^2)^3_{}\,,
\end{eqnarray}
and
\begin{eqnarray}
\Gamma_{\chi_2^{}\rightarrow
\chi_1^{}Z}^{}&=&\Gamma_{\bar{\chi}_2^{}\rightarrow
\bar{\chi}_1^{}Z}^{}\nonumber\\
&\simeq&
\frac{c_R^{2}s_W^{2}}{2\pi}|\lambda_{12}^{}|^2m_{\chi_2^{}}^3
F_2^{}\left(\frac{m_{\chi_1^{}}^2}{m_{\chi_2^{}}^{2}},\frac{m_{Z}^2}{m_{\chi_2^{}}^{2}}\right)\,,\nonumber\\
\Gamma_{\chi_2^{}\rightarrow
\chi_1^{}Z_X^{}}^{}&=&\Gamma_{\bar{\chi}_2^{}\rightarrow
\bar{\chi}_1^{}Z_X^{}}^{}\nonumber\\
&\simeq& \frac{1}{2\pi}|\kappa_{12}^{}|^2m_{\chi_2^{}}^3
F_2^{}\left(\frac{m_{\chi_1^{}}^2}{m_{\chi_2^{}}^{2}},\frac{m_{Z_X^{}}^2}{m_{\chi_2^{}}^{2}}\right)~
~\textrm{with}\nonumber\\
&&
F_2^{}(x,y)=\left[(1-x)^2_{}-\frac{1}{2}y(1+x+y)\right]\nonumber\\
&&\quad\quad\quad\quad\quad\times \sqrt{(1-x-y)^2-4xy}\,.
\end{eqnarray}

\begin{figure}
\vspace{9.0cm} \epsfig{file=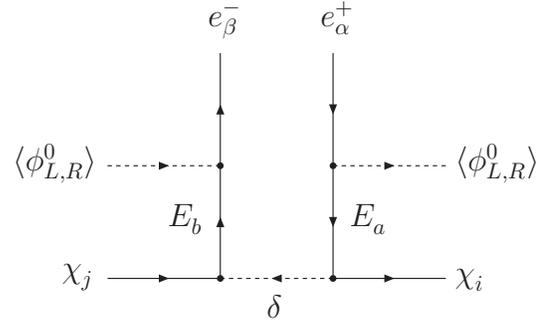, bbllx=3cm, bblly=6.0cm,
bburx=13cm, bbury=16cm, width=8.5cm, height=8.5cm, angle=0, clip=0}
\vspace{-12.7cm} \caption{\label{treedecay} The heavier dark matter
fermion $\chi_{j}^{}(\bar{\chi}_i^{})$ decays into the lighter dark
matter fermion $\chi_{i}^{}(\bar{\chi}_j^{})$ and a charged lepton
pair $e_{\alpha}^{+}e_\beta^{-}$.}
\end{figure}

In Fig. \ref{treedecay}, we see the heavier dark matter fermion
$\chi_2^{}$ can also decay into the lighter dark matter fermion
$\chi_1^{}$ and a pair of the SM charged leptons $e_\alpha^+
e_\beta^-$ at tree level. These three-body decay modes have the
decay widths as below,
\begin{eqnarray}
\Gamma_{\chi_2^{}\rightarrow \chi_1^{}l^+_{}
l_{}^-}^{}&=&\Gamma_{\bar{\chi}_2^{}\rightarrow
\bar{\chi}_1^{}l^+_{}
l_{}^-}^{}=\sum_{\alpha,\beta}^{}\Gamma_{\chi_2^{}\rightarrow
\chi_1^{}e_\alpha^+
e_\beta^-}^{}\nonumber\\
&\simeq&\frac{1}{2^{9}_{}\pi^3_{}}(R_{11}^{}R_{22}^{}+R_{11}^{}L_{22}^{}
+R_{22}^{}L_{11}^{}\nonumber\\
&&+L_{11}^{}L_{22}^{})\frac{m_2^5}{M_\delta^4}F_3^{}
\left(\frac{m_{\chi_1^{}}^2}{m_{\chi_2^{}}^2}\right)\,,
\end{eqnarray}
with
\begin{eqnarray}
R_{ii}^{}&=&\left(f_L^\dagger\frac{v_R^{}}{\sqrt{2}M_E^{}}y_{E}^{R\dagger}
y_{E}^R\frac{v_R^{}}{\sqrt{2}M_E^{}}f_L^{}\right)_{ii}^{}\,,\nonumber\\
L_{ii}^{}&=&\left(f_R^\dagger\frac{v_L^{}}{\sqrt{2}M_E^{}}y_{E}^{L\dagger}
y_{E}^L\frac{v_L^{}}{\sqrt{2}M_E^{}}f_R^{}\right)_{ii}^{}\,,
\end{eqnarray}
and
\begin{eqnarray}
F_3^{}(x)=\frac{1}{12}(1-8x+8x^3_{}-x^4_{})-x^2_{}\ln x\,.
\end{eqnarray}

The heavier dark matter fermion $\chi_2^{}$ can have a very long
lifetime. For example, by inputting
\begin{eqnarray}
&&g_{B-L}^{}=0.428\,,~~g_X^{}=0.592\,,~~\epsilon=10^{-7}_{}\,,\nonumber\\
&&f_L^{}=f_R^{}=\textrm{diag}\left\{\sqrt{\frac{\sqrt{2}m_e^{}}{v_L^{}}},
\sqrt{\frac{\sqrt{2}m_\mu^{}}{v_L^{}}},\sqrt{\frac{\sqrt{2}m_\tau^{}}{v_L^{}}}\right\}\,,\nonumber\\
&&M_{E_{1,2,3}^{}}^{}=\frac{1}{\sqrt{2}}v_R^{}=900\,\textrm{TeV}\,,~~M_\delta^{}=
10^{16}_{}\,\textrm{GeV}\,,\nonumber\\
&&m_{\chi_2^{}}^{}=262\,\textrm{GeV}\,,~~
m_{\chi_1^{}}^{}=20\,\textrm{GeV}\,,
\end{eqnarray}
we can obtain
\begin{eqnarray}
\Gamma_{\chi_2^{}\rightarrow
\chi_1^{}\gamma}^{}&=&\Gamma_{\bar{\chi}_2^{}\rightarrow
\bar{\chi}_1^{}\gamma}^{}=5.03\times 10^{-53}_{}\,\textrm{GeV}\,,\nonumber\\
\Gamma_{\chi_2^{}\rightarrow
\chi_1^{}Z}^{}&=&\Gamma_{\bar{\chi}_2^{}\rightarrow
\bar{\chi}_1^{}Z}^{}
=1.23\times 10^{-53}_{}\,\textrm{GeV}\,,\nonumber\\
\Gamma_{\chi_2^{}\rightarrow
\chi_1^{}Z_X^{}}^{}&=&\Gamma_{\bar{\chi}_2^{}\rightarrow
\bar{\chi}_1^{}Z_X^{}}^{}
=2.22\times 10^{-57}_{}\,\textrm{GeV}\,,\nonumber\\
\Gamma_{\chi_2^{}\rightarrow \chi_1^{}l^+_{}
l_{}^-}^{}&=&\Gamma_{\bar{\chi}_2^{}\rightarrow
\bar{\chi}_1^{}l^+_{}l^-_{}}^{}\nonumber\\
&=&6.20\times 10^{-66}_{}\,\textrm{GeV}\,,
\end{eqnarray}
and then
\begin{eqnarray}
\textrm{Br}_{\chi_1^{}\gamma}^{}&=&\textrm{Br}(\chi_2^{}\rightarrow
\chi_1^{}\gamma)+\textrm{Br}(\bar{\chi}_2^{}\rightarrow
\bar{\chi}_1^{}\gamma)\nonumber\\
&=&\frac{\Gamma_{\chi_2^{}\rightarrow
\chi_1^{}\gamma}^{}+\Gamma_{\bar{\chi}_2^{}\rightarrow
\bar{\chi}_1^{}\gamma}^{}}{\Gamma_{\chi_2^{}}^{}+\Gamma_{\bar{\chi}_2^{}}^{}}\simeq
80.3\%\,,\nonumber\\
\tau_{\chi_2^{}}^{}&=&\tau_{\bar{\chi}_2^{}}^{}=\frac{1}{\Gamma_{\chi_2^{}}^{}}\simeq
1.05\times 10^{28}_{}\,\textrm{sec}\,,
\end{eqnarray}
with $\Gamma_{\chi_2^{}}^{}$ and $\Gamma_{\bar{\chi}_2^{}}^{}$ being
the total decay width,
\begin{eqnarray}
\Gamma_{\chi_2^{}}^{}=
\Gamma_{\bar{\chi}_2^{}}^{}&=&\Gamma_{\chi_2^{}\rightarrow
\chi_1^{}\gamma}^{}+\Gamma_{\chi_2^{}\rightarrow \chi_1^{}Z}^{}
+\Gamma_{\chi_2^{}\rightarrow \chi_1^{}Z_X^{}}^{}\nonumber\\
&&
+\Gamma_{\chi_2^{}\rightarrow \chi_1^{}l^+_{} l_{}^-}^{}\nonumber\\
&\simeq&\Gamma_{\chi_2^{}\rightarrow
\chi_1^{}\gamma}^{}+\Gamma_{\chi_2^{}\rightarrow
\chi_1^{}Z}^{}\nonumber\\
&=&6.26\times 10^{-53}_{}\,\textrm{GeV}\,,
\end{eqnarray}
Furthermore, the photons from the decays $\chi_2^{}\rightarrow
\chi_1^{}\gamma$ and $\bar{\chi}_2^{}\rightarrow
\bar{\chi}_1^{}\gamma$ have the determined energy
\begin{eqnarray}
\label{menergy}E_\gamma^{}&=&\frac{m_{\chi_2^{}}^{2}
-m_{\chi_1^{}}^2} {2m_{\chi_2^{}}^{}}\simeq 130\,\textrm{GeV}\,.
\end{eqnarray}

The gamma-ray flux from the dark matter decays $\chi_2^{}\rightarrow
\chi_1^{}\gamma$ and $\bar{\chi}_2^{}\rightarrow
\bar{\chi}_1^{}\gamma$ can be written as
\begin{eqnarray}
\label{mflux}
\frac{d\Phi}{dEd\Omega}&=&\frac{\Omega_{\chi_{2}^{}+\bar{\chi}_2^{}}^{}h^2_{}}{\Omega_{\textrm{DM}}^{}h^2_{}}
\frac{\textrm{Br}_{\chi_1^{}\gamma}^{}}{4\pi
m_{\chi_2^{}}^{}\tau_{\chi_2^{}}^{}}\int_{l.o.s}^{}
ds\rho[r(s,\psi)]\frac{dN}{dE}\,,\nonumber\\
&&
\end{eqnarray}
where $dN/dE$ is the differential gamma spectrum per dark matter
decay with $E$ being the gamma-ray energy,
$r(s,\psi)=(r_{\odot}^2+s^2_{}-2r_{\odot}^{}s\cos\psi)^{1/2}_{}$ is
the coordinate centered on the GC with $s$ being the distance from
the Sun along the line-of-sight (l.o.s), $r_{\odot}^{}$ being the
distance from the Sun to the GC and $\psi$ being the angle between
the direction of observation in the sky and the GC,
$\rho[r(s,\psi)]$ is the dark matter density profile. In Ref.
\cite{bg2012}, the authors have shown that the decay of a
single-component dark matter fermion into a neutrino and a photon
can explain the gamma-ray line in the Fermi-LAT data \cite{bg2012},
i.e.
\begin{eqnarray}
\label{sflux} \frac{d\Phi}{dEd\Omega}&=&
\frac{\textrm{Br}_{\nu^{}\gamma}^{}}{4\pi
m_{\chi}^{}\tau_{\chi}^{}}\int_{l.o.s}^{}
ds\rho[r(s,\psi)]\frac{dN}{dE}\,.
\end{eqnarray}
Here
\begin{eqnarray}
\label{senergy} m_\chi^{}\simeq 2E_\gamma^{}\simeq
260\,\textrm{GeV}\,,
\end{eqnarray}
is the dark matter mass, $\tau_\chi^{}$ is the dark matter lifetime,
$\textrm{Br}_{\nu\gamma}^{}$ is the branching ratio of the decay
modes $\chi\rightarrow \nu\gamma$. By comparing Eqs.
(\ref{menergy}-\ref{mflux}) and (\ref{sflux}-\ref{senergy}), we can
take
\begin{eqnarray}
&&\frac{m_{\chi_2^{}}^{2} -m_{\chi_1^{}}^2}
{2m_{\chi_2^{}}^{}}=m_{\chi}^{}\,,
~\textrm{Br}_{\chi_1^{}\gamma}^{}=\textrm{Br}_{\nu\gamma}^{}\,,\nonumber\\
&&\tau_{\chi_2^{}}^{}=\frac{\Omega_{\chi_{2}^{}+\bar{\chi}_2^{}}^{}h^2_{}}{\Omega_{\textrm{DM}}^{}h^2_{}}
\frac{m_{\chi}^{}}{m_{\chi_2^{}}^{}}\tau_{\chi}^{}\,,
\end{eqnarray}
to account for the fitting results \cite{bg2012}. For example, we
can obtain the Fermi-LAT gamma-ray line by inputting
\begin{eqnarray}
&&m_{\chi_2^{}}^{}=262\,\textrm{GeV}\,,~~
m_{\chi_1^{}}^{}=20\,\textrm{GeV}\,,\nonumber\\
&&
\frac{\Omega_{\chi_{2}^{}+\bar{\chi}_2^{}}^{}h^2_{}}{\Omega_{\textrm{DM}}^{}h^2_{}}=99.1\%\,,~~
\textrm{Br}_{\chi_1^{}\gamma}^{}=
80.3\%\,,\nonumber\\
&&\tau_{\chi_2^{}}^{}= 1.05\times 10^{28}_{}\,\textrm{sec}\,.
\end{eqnarray}

\section{Dark matter scattering and self-interaction}

\begin{figure}
\vspace{5.5cm} \epsfig{file=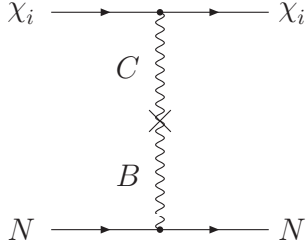, bbllx=2.0cm, bblly=6.0cm,
bburx=12.0cm, bbury=16cm, width=8.2cm, height=8.2cm, angle=0,
clip=0} \vspace{-9.5cm} \caption{\label{scattering} The dark matter
scattering $\chi_i^{}N\rightarrow \chi_i^{}N$. Here $\chi_{i}^{}$
and $N$ denote the dark matter fermions and the ordinary nucleons
while $B$ and $C$ are the $U(1)_{B-L}^{}$ and $U(1)_X^{}$ gauge
fields. For simplicity, we don't show the scattering
$\bar{\chi}_i^{}N\rightarrow \bar{\chi}_i^{}N$.}
\end{figure}

As shown in Fig. \ref{scattering}, the dark matter fermions
$\chi_{1,2}^{}$ can scatter off the nucleon $N$ through the kinetic
mixing between the $U(1)_X^{}$ and $U(1)_{B-L}^{}$ gauge fields. The
elastic scattering cross section is computed by
\begin{eqnarray}
\sigma_{\chi_i^{}N}^{}&=&\sigma_{\chi_i^{}N\rightarrow
\chi_i^{}N}^{}=\sigma_{\bar{\chi}_i^{}N\rightarrow
\bar{\chi}_i^{}N}^{}\simeq\frac{\epsilon^2_{}g_X^2
g_{B-L}^2}{324\pi}\frac{\mu^2_{r}}{m_{Z_X^{}}^4}\nonumber\\
&=&3.9\times
10^{-45}_{}\,\textrm{cm}^2_{}\left(\frac{\epsilon}{10^{-7}_{}}\right)^2_{}
\left(\frac{g_X^{}}{0.592}\right)^2_{}\nonumber\\
&&\times\left(\frac{g_{B-L}^{}}{0.428}\right)^2_{}
\left(\frac{500\,\textrm{MeV}}{m_{Z_X^{}}^{}}\right)^4_{}\left(\frac{\mu_r^{}}{1\,
\textrm{GeV}}\right)^2_{}\,.
\end{eqnarray}
Here we have defined the reduced mass,
\begin{eqnarray}
\mu_r^{}=\frac{m_{\chi_i^{}}^{}m_N^{}}
{m_{\chi_i^{}}^{}+m_N^{}}\simeq m_N^{}\simeq
1\,\textrm{GeV}~~\textrm{for}~~m_{\chi_i^{}}^{}\gg m_N^{}\,.
\end{eqnarray}
Such scattering can be measured by the dark matter direct detection
experiments. The event rate per unit time per nucleon should be
\begin{eqnarray}
R_{\chi_i^{}}^{}\approx\frac{\Omega_{\chi_{i}^{}+\bar{\chi}_i^{}}^{}h^2_{}}{\Omega_{\textrm{DM}}^{}h^2_{}}
\frac{\rho_{\odot}^{}}{m_{\chi_i^{}}^{}}\sigma_{\chi_i^{} N}^{}
\end{eqnarray}
with $\rho_{\odot}^{}$ being the local dark matter density.
Currently, the measured experimental rate is given in the
single-component dark matter hypothesis,
\begin{eqnarray}
R_{\chi}^{}\approx \frac{\rho_{\odot}^{}}{m_{\chi}^{}}\sigma_{\chi
N}^{}
\end{eqnarray}
where $\sigma_{\chi N}^{}$ denotes the cross section of a
single-component dark matter $\chi$ with the mass $m_\chi^{}$
scattering off the nucleon $N$. The XENON10 and XENON100 experiments
\cite{angle2011,aprile2012} have stringently put an upper bound
$\sigma_{\chi N}^{\textrm{exp}}$ on the cross section $\sigma_{\chi
N}^{}$ for a given mass $m_\chi^{}$. Therefore, we should constrain
\begin{eqnarray}
\sigma_{\chi_i^{}
N}^{}<\frac{\Omega_{\textrm{DM}}^{}h^2_{}}{\Omega_{\chi_{i}^{}+\bar{\chi}_i^{}}^{}h^2_{}}\sigma_{\chi
N}^{\textrm{exp}}\,.
\end{eqnarray}
For a proper parameter choice such as
\begin{eqnarray}
&&m_{\chi_1^{}}^{}=20\,\textrm{GeV}\,,~m_{\chi_2^{}}^{}=262\,\textrm{GeV}\,,\nonumber\\
&&~\sigma_{\chi_1^{}N}^{}=\sigma_{\chi_2^{}N}^{}=3.9\times
10^{-45}_{}\,\textrm{cm}^2_{}\,,\nonumber\\
&&\frac{\Omega_{\chi_{1}^{}+\bar{\chi}_1^{}}^{}h^2_{}}{\Omega_{\textrm{DM}}^{}h^2_{}}\simeq
0.9\%\,,~ \frac{\Omega_{\chi_{2}^{}+\bar{\chi}_2^{}}^{}h^2_{}}
{\Omega_{\textrm{DM}}^{}h^2_{}}\simeq99.1\%\,,
\end{eqnarray}
the dark matter fermions $\chi_{1,2}^{}$ can be verified by the
ongoing and forthcoming dark matter direct detection experiments.

\begin{figure*}
\vspace{5.5cm} \epsfig{file=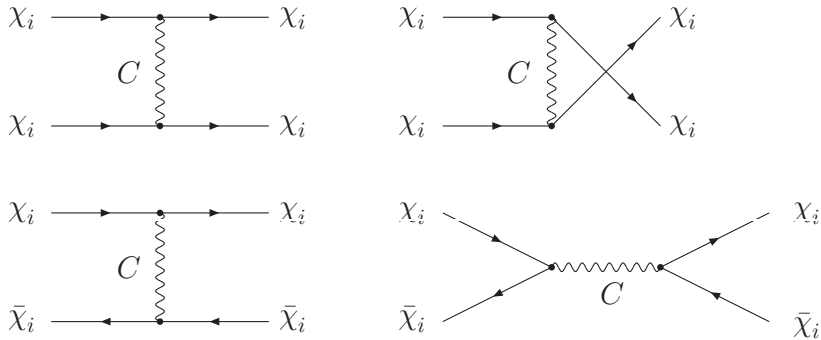, bbllx=3.5cm, bblly=6.0cm,
bburx=13.5cm, bbury=16cm, width=8.2cm, height=8.2cm, angle=0,
clip=0} \vspace{-8cm} \caption{\label{self} The dark matter
self-interactions $\chi_i^{}\chi_i^{}\rightarrow\chi_i^{}\chi_i^{}$
and $\chi_i^{}\bar{\chi}_i^{}\rightarrow\chi_i^{}\bar{\chi}_i^{}$.
Here $\chi_{i}^{}$ denotes the dark matter fermions while $C$ is the
$U(1)_X^{}$ gauge filed. For simplicity, we don't show the processes
$\bar{\chi}_i^{}\bar{\chi}_i^{}\rightarrow\bar{\chi}_i^{}\bar{\chi}_i^{}$.}
\end{figure*}

Furthermore, the dark matter fermions $\chi_{1,2}^{}$ can have a
self-interaction as shown in Fig. \ref{self}. For
$m_{\chi_{i}^{}}^{}\gg m_{Z_X^{}}^{}$, the self-interacting cross
section should be

\begin{eqnarray}
\frac{\sigma_{\chi_i^{}\chi_i^{}\rightarrow
\chi_i^{}\chi_i^{}}^{}}{m_{\chi_i^{}}^{}}
&=&\frac{\sigma_{\bar{\chi}_i^{}\bar{\chi}_i^{}\rightarrow
\bar{\chi}_i^{}\bar{\chi}_i^{}}^{}}{m_{\chi_i^{}}^{}}
\simeq\frac{g_X^4}{628\pi}\frac{m_{\chi_i^{}}^{}}{m_{Z_X^{}}^4}\,,\nonumber\\
\frac{\sigma_{\chi_i^{}\bar{\chi}_i^{}\rightarrow
\chi_i^{}\bar{\chi}_i^{}}^{}}{m_{\chi_i^{}}^{}}
&\simeq&\frac{g_X^4}{324\pi}\frac{m_{\chi_i^{}}^{}}{m_{Z_X^{}}^4}\,,
\end{eqnarray}
from which we read

\begin{eqnarray}
\label{selfcs} \frac{\sigma_{\chi_1^{}\bar{\chi}_1^{}\rightarrow
\chi_1^{}\bar{\chi}_1^{}}^{}}{m_{\chi_1^{}}^{}}&\simeq&2\frac{\sigma_{\chi_1^{}\chi_1^{}\rightarrow
\chi_1^{}\chi_1^{}}^{}}{m_{\chi_1^{}}^{}}=2\frac{\sigma_{\bar{\chi}_1^{}\bar{\chi}_1^{}\rightarrow
\bar{\chi}_1^{}\bar{\chi}_1^{}}^{}}{m_{\chi_1^{}}^{}}\nonumber\\
&\simeq&3.0\times
10^{-43}_{}\,\textrm{cm}^3_{}\left(\frac{g_X^{}}{0.592}\right)^4_{}
\nonumber\\
&&\times\left(\frac{m_{\chi_1^{}}^{}}{20\,\textrm{GeV}}\right)
\left(\frac{500\,\textrm{MeV}}{m_{Z_X^{}}^{}}\right)^4_{}\,,\nonumber\\
\frac{\sigma_{\chi_2^{}\bar{\chi}_2^{}\rightarrow
\chi_2^{}\bar{\chi}_2^{}}^{}}{m_{\chi_2^{}}^{}}&\simeq&2\frac{\sigma_{\chi_2^{}\chi_2^{}\rightarrow
\chi_2^{}\chi_2^{}}^{}}{m_{\chi_2^{}}^{}}=2\frac{\sigma_{\bar{\chi}_2^{}\bar{\chi}_2^{}\rightarrow
\bar{\chi}_2^{}\bar{\chi}_2^{}}^{}}{m_{\chi_2^{}}^{}}\nonumber\\
&\simeq&3.9\times
10^{-42}_{}\,\textrm{cm}^3_{}\left(\frac{g_X^{}}{0.592}\right)^4_{}
\nonumber\\
&&\times\left(\frac{m_{\chi_2^{}}^{}}{262\,\textrm{GeV}}\right)
\left(\frac{500\,\textrm{MeV}}{m_{Z_X^{}}^{}}\right)^4_{}\,.
\end{eqnarray}
For a single-component dark matter with the mass $m$, its
self-interacting cross section $\sigma$ has an upper bound
$\sigma/m<4.4\times 10^{-42}_{}\,\textrm{cm}^3_{}$ \cite{rmcgb2007}.
We hence should require
\begin{eqnarray}
\label{selfbound}
\frac{\Omega_{\chi_i^{}}^{}h^2}{\Omega_{\textrm{DM}}^{}h^2}\frac{\sigma_{\chi_i^{}\chi_i^{}\rightarrow
\chi_i^{}\chi_i^{}}^{}}{m_{\chi_i^{}}^{}}&=&
\frac{\Omega_{\chi_i^{}+\bar{\chi}_i^{}}^{}h^2}{2\Omega_{\textrm{DM}}^{}h^2}\frac{\sigma_{\chi_i^{}\chi_i^{}\rightarrow
\chi_i^{}\chi_i^{}}^{}}{m_{\chi_i^{}}^{}}\,,\nonumber\\
\frac{\Omega_{\bar{\chi}_i^{}}^{}h^2}{\Omega_{\textrm{DM}}^{}h^2}\frac{\sigma_{\bar{\chi}_i^{}\bar{\chi}_i^{}\rightarrow
\bar{\chi}_i^{}\bar{\chi}_i^{}}^{}}{m_{\chi_i^{}}^{}}&=&
\frac{\Omega_{\chi_i^{}+\bar{\chi}_i^{}}^{}h^2}{2\Omega_{\textrm{DM}}^{}h^2}\frac{\sigma_{\bar{\chi}_i^{}\bar{\chi}_i^{}\rightarrow
\bar{\chi}_i^{}\bar{\chi}_i^{}}^{}}{m_{\chi_i^{}}^{}}\,,\nonumber\nonumber\\
\frac{\Omega_{\chi_i^{}+\bar{\chi}_i^{}}^{}h^2}
{\Omega_{\textrm{DM}}^{}h^2}\frac{\sigma_{\chi_i^{}\bar{\chi}_i^{}\rightarrow
\chi_i^{}\bar{\chi}_i^{}}^{}}{m_{\chi_i^{}}^{}}&<&4.4\times
10^{-42}_{}\,\textrm{cm}^3_{}\,.
\end{eqnarray}
Clearly, the parameter choice (\ref{selfcs}) can satisfy the limit
(\ref{selfbound}) since we have
$\frac{\Omega_{\chi_i^{}+\bar{\chi}_i^{}}^{}h^2}
{\Omega_{\textrm{DM}}^{}h^2}<1$.

\section{Summary}

It was suggested that the $130\,\textrm{GeV}$ gamma-ray line hinted
by the Fermi-LAT data could be understood by the dark matter
annihilation or decay into monochromatic photons. We hence propose a
multi-component dark matter model, where a heavier dark matter
fermion mostly decays into a lighter dark matter fermion and a
photon, to explain the Fermi-LAT gamma-ray line. In our model, the
neutral dark matter fermions have a highly suppressed magnetic
moment at one-loop level because of their Yukawa couplings to a
charged scalar and three non-SM leptons. The new scalar besides the
dark matter fermions is gauged by a $U(1)_X^{}$ symmetry which will
be spontaneously broken below the GeV scale. As for the non-SM
leptons, they play an essential role for generating the SM lepton
masses in the $SU(3)_c^{}\times SU(2)_L^{}\times SU(2)_R^{}\times
U(1)_{B-L}^{}$ left-right symmetric models for the universal seesaw
scenario where the strong CP problem can be solved without an axion.
The dark matter fermions can obtain a thermally produced relic
density through their annihilations into the $U(1)_X^{}$ gauge and
Higgs fields. The kinetic mixing between the $U(1)_X^{}$ and
$U(1)_{B-L}^{}$ gauge fields can result in a testable dark matter
scattering.

\end{document}